 \documentclass{menu97}    % US letter paper size
 \usepackage{epsfig}

\begin{document}
    \setlength{\baselineskip}{2.6ex}
\begin{flushright}
HIP - 1997 - 49 / TH
\end{flushright}

\title{Low-Energy Pion-Nucleon Interaction and the Sigma-Term\thanks{Invited
talk presented at the ``7th International Symposium on Meson-Nucleon
Physics and the Structure of the Nucleon'', Vancouver, B.C., Canada, July 28-
Aug. 1, 1997.}}
\author{M.E. Sainio\\
{\em Department of Physics, University of Helsinki, P.O. Box 9,
00014 Helsinki, Finland}}

\maketitle

\begin{abstract}
\setlength{\baselineskip}{2.6ex}
A dispersion framework, appropriate for discussing the low-energy pion-nucleon 
interaction, is reviewed. Sensitivity of the isoscalar $D$-amplitude at
the unphysical Cheng-Dashen point to input from
different phase shift analyses and low-energy experiments is discussed.
\end{abstract}

\setlength{\baselineskip}{2.6ex}

\section*{INTRODUCTION}
In the limit of vanishing $u$- and $d$-quark masses the QCD Hamiltonian
has chiral $SU(2) \times SU(2)$ symmetry. The symmetry group would be
$SU(3) \times SU(3)$, if also the strange quark is treated as massless.
The QCD vacuum does not, however, have the same symmetry, and the symmetry
is said to be hidden, or spontaneously broken. As a consequence, the Goldstone 
theorem dictates that massless pseudoscalar mesons appear. 
They acquire a mass through
the small, but nonzero, quark masses. The quark masses also shift the mass
of the proton. The proton matrix element of the light quark mass
term in the Hamiltonian, the sigma-term, is a measure of the explicit chiral 
symmetry breaking. 

\section*{THE SIGMA TERM}

The sigma-term is defined as
\begin{eqnarray}
\sigma=\frac{\hat{m}}{2 m} \langle p | \bar{u}u + \bar{d}d |p\rangle,
\end{eqnarray}
where $\hat{m} =\frac{1}{2}(m_u + m_d)$, $m$ is the proton mass and
$|p\rangle$ denotes the physical one-proton state normalized as
$\langle p' |p\rangle = (2 \pi)^3 \, 2 p_0 \, \delta(\bf{p'}-\bf{p})$.
Algebraically $\sigma$ can be written in the form
\begin{eqnarray}
\sigma=\frac{\hat{m}}{2 m} \frac{\langle p | \bar{u}u + \bar{d}d - 2 \bar{s}s 
|p\rangle}{1-y},
\end{eqnarray}
where the parameter $y$, the strange quark content of the proton, is defined
as
\begin{eqnarray}
y=\frac{2 \langle p | \bar{s}s |p\rangle}
{\langle p | \bar{u}u + \bar{d}d |p\rangle}.
\end{eqnarray}
The combination $\sigma (1-y)$ can be calculated in Chiral Perturbation
Theory (CHPT), and with $m_s/\hat{m} \simeq 25$ an
estimate can be given in different orders of the quark mass expansion
\begin{eqnarray}
\sigma (1-y) &\simeq& 26 \; \; {\rm MeV} \hspace{2.7cm} {\cal O}(m_q)
             \nonumber \\
             &\simeq& 35 \pm 5 \; \; {\rm MeV} \hspace{2.cm}{\cal O}(m_q^{3/2})
             \nonumber \\
             &\simeq& 36 \pm 7 \; \; {\rm MeV} \hspace{2.cm}{\cal O}(m_q^2).
\end{eqnarray}
The corrections to the leading order calculation reflect the meson cloud
effects. The result of the order $m_q^{3/2}$ is from Ref. [1,2] and the  
$m_q^2$ result from Ref. [3]. 
In the OZI rule limit ($y=0$) a value for
$\sigma$ follows to a  given order of the quark mass expansion.

A low-energy theorem of chiral symmetry allows for another determination
of the $\sigma$-term. Namely, the pion-nucleon scattering amplitude at a 
particular unphysical (but on-shell) point, so called the Cheng-Dashen point,
can be related to the nucleon scalar form factor up to corrections of
order $\mu^4 \log \mu^2$ [4]
\begin{eqnarray}
\Sigma \equiv F_\pi^2 \bar{D}^+(\nu=0,t=2 \mu^2) = \sigma(2 \mu^2)+\Delta_R.
\end{eqnarray}
Here $F_\pi = 92.4$ MeV is the pion decay constant, $\bar{D}^+$ denotes
the isoscalar combination of the pion-nucleon amplitudes $D = A + \nu B$,
where the pseudovector Born term has been subtracted, i.e.
$\bar{D} = D - D_{pv}$, $\nu=(s-u)/4m$
in terms of the Mandelstam variables and $\mu$ is the pion mass.
The scalar form factor $\sigma(t)$ is defined by
\begin{eqnarray}
\bar{u}(p')\sigma(t)u(p)=\langle p' |\hat{m}( \bar{u}u + \bar{d}d) |p\rangle
\; \; ; \; \; t=(p'-p)^2.
\end{eqnarray}
The correction term, $\Delta_R$, is formally of order $\mu^4$, and
to one loop in chiral perturbation theory [5] $\Delta_R = 0.35$ MeV.
An estimate for the upper limit of $\Delta_R$ including all terms of the 
order $m_q^2$ is 2 MeV [6] (it is the upper limit, because the sign of the 
contribution from the scalar meson is unknown; the conservative choice is 
taken in [6]). Also, to this order in the quark mass expansion no chiral
logarithm is found.
In view of this result, the approximation
$\Sigma \simeq \sigma(2 \mu^2)$
is good enough for the present level of accuracy. In addition to fixing
the value of $\Sigma$ from pion-nucleon information, one still has to 
consider the
$t$-dependence of the scalar form factor to get an estimate for 
$\sigma$ ($=\sigma(t=0))$, so the quantity of interest is
\begin{eqnarray}
\Delta_\sigma=\sigma(2 \mu^2)-\sigma(0).
\end{eqnarray}
The first calculation was published in Ref. [7] and later more refined
results were given in [1,5]. CHPT to one loop gives [5]
$\Delta_\sigma \simeq 5 \; \; {\rm MeV}$,
which, however, corresponds to the treatment of the pion-pion scattering
at tree level.
In [8] $\Delta_\sigma$ was calculated with a dispersion relation with 
$\pi \pi$ input consistent with CHPT. The result is
\begin{eqnarray}
\Delta_\sigma = 15.2 \pm 0.4 \; \; {\rm MeV},
\end{eqnarray}
which is also obtained by Bernard et al. [9] in a calculation which
includes some of the contributions of order $q^4$.

Using Eqs. 5 and 7 the $\sigma$ can be calculated once $\Sigma$ is
fixed from pion-nucleon information, namely
\begin{eqnarray}
\sigma=\Sigma -(\Delta_\sigma+\Delta_R),
\end{eqnarray}
i.e. the difference between $\Sigma$ and $\sigma$ is about 15 MeV with
relatively small uncertainty in comparison with the current uncertainties
in $\Sigma$ and the CHPT result for $\sigma(1-y)$.

\section*{DETERMINATION OF $\Sigma$}

Some time ago Koch used [10] hyperbolic dispersion relations to determine 
$\Sigma$ and got the result
\begin{eqnarray}
\Sigma = 64 \pm 8 \; \; {\rm MeV},
\end{eqnarray}
where the uncertainty reflects the internal consistency of the method.
To be able to estimate the uncertainty due to the experimental errors
of the low-energy data, a dispersion method was proposed in Ref. [11]. 
That approach involves six forward dispersion relations, 
the standard ones for $D^{\pm}$ and $B^{\pm}$ and, in addition, two
dispersion relations for the newly defined amplitudes
\begin{eqnarray}
E^{\pm}(\omega) = \frac{\partial}{\partial t} (A^{\pm} + 
\omega B^{\pm})|_{t=0},
\end{eqnarray}
where $\omega$ is the pion total energy.
The point here is that in the low-energy range the 2 $s$-waves and 4 $p$-waves
matter, and the $d$- and $f$-waves can be treated as corrections 
(to be taken from
an existing phase shift analysis; KA85). The six dispersion relations will
give six energy dependent functions, the partial wave amplitudes $(l=0,1)$. 
The method is a
low-energy approach; it makes sense only to the extent that $d$- and higher
partial waves can be treated as corrections. In Ref. [12] it was assumed
that the $d$-waves were accurate to 30 \%, and the error estimate there
involved variations of the $d$-waves by that amount around the KA85 values 
[13].
In addition to the low-energy $l \geq 2$ waves, input is needed for
the higher energies. (The ``high'' energy piece starts here at the pion
momentum $k_0= 185$ MeV/c corresponding to a laboratory kinetic energy
of 92 MeV.) In this range the imaginary parts of the invariant amplitudes
$D^\pm$, $B^\pm$ and $E^\pm$ are needed, and they are constructed from
partial wave solutions of the Karlsruhe group (KH80, KA84 and KA85) and,
also, from solutions FA93, SM95 [14] and SM97 of the VPI group.
For the asymptotic behaviour of the amplitudes Karlsruhe forms [15] have
been adopted.

The six dispersion relations fully fix the amplitudes up to two 
subtraction constants both of which have their origin in the $A^+$ amplitude. 
The subtraction constant
in the dispersion relation for $D^+$ is proportional to the $s$-wave
scattering length $a^+_{0+}$, and the subtraction constant for the $E^+$
amplitude to $a^+_{1+}$. Once these two threshold parameters are known, the
amplitude is fully fixed. Experimental data at low energies (below the
momentum $k_0$) will be needed to determine $a^+_{0+}$ and $a^+_{1+}$.
At the lowest energies there are results from differential cross section
measurements which are used as input in the $\chi^2$ search.
For each data set the contribution to the total
$\chi^2$ is
\begin{eqnarray}
\chi^{2} \equiv \left( {z-1 \over \Delta z} \right)^{2} + 
          \sum_{{\rm DATA}} \left( {z \sigma^{*} (\theta)-\sigma (\theta) \over
                \Delta \sigma (\theta)} \right)^{2},
\end{eqnarray}
where $\sigma^{*}(\theta)$ is the experimental differential cross section
with error $\Delta \sigma (\theta)$. The computed cross section is denoted 
by $\sigma (\theta)$, and it contains the full observable cross section
constructed from the solutions for the hadronic amplitudes
from the dispersion relations, and the electromagnetic corrections according
to the formalism of Tromborg et al. [16]. Also, the P$_{33}$ splitting
between the $\pi^- p$ and $\pi^+ p$ channels has been taken into account.
For each data set there is
one parameter $z$ which takes care of the normalization of that set, i.e.
in the present analysis there is a contribution to the $\chi^2$ from the
normalization. However, it does not depend on the number of data points in
the data set. The experimental normalization uncertainty has been used
for $\Delta z$, if given in the experimental paper. The iterative procedure
for solving the coupled dispersion relations has then the following steps:
\begin{enumerate}
\item Take initial values for $(a^+_{0+}, a^+_{1+})$, and set 
      $\delta_l \equiv 0$ for $l\leq1$ and $k_{\rm lab} < k_0$.
\item Construction of the imaginary parts of the six invariant amplitudes for 
      all momenta.
\item Solution of the dispersion relations.
\item Partial waves from the real parts of the invariant amplitudes at low
      energy.
\item Back to step 2 until the phase shifts stabilize.
\end{enumerate}
Usually only a few iterations are needed to get smooth partial wave amplitudes.

It is practical to write the subthreshold expansion at the origin of the
$(\nu,t)$ -plane
\begin{eqnarray}
\bar{D}^+=d^+_{00}+d^+_{10}\, \nu^2+d^+_{01}\, t+d^+_{20} \,\nu^4+
          d^+_{11}\, \nu^2 t + ...,
\end{eqnarray}
where two of the constants can be directly related to the dispersion relations
\begin{eqnarray}
d^+_{00}=\bar{D}^+(0), \; \; \; \; d^+_{01}=\bar{E}^+(0).
\end{eqnarray}
The expression for $\Sigma$ then gets the form
\begin{eqnarray}
\Sigma=F^2_\pi (d^+_{00} + 2 \mu^2 d^+_{01}) + \Delta_D \equiv \Sigma_d +
       \Delta_D,
\end{eqnarray}
where the curvature term $\Delta_D$ is dominated by the $\pi \pi$ cut 
giving [8]
\begin{eqnarray} 
\Delta_D = 11.9 \pm 0.6 \; \; {\rm MeV}.
\end{eqnarray}
However, there is also the left-hand cut contribution, so a new 
$\pi N$ analysis
could modify this number slightly. The term linear in $t$ is a 
relatively
sensitive quantity as can be seen from the numbers for the solutions A and B 
in [12]
\begin{eqnarray}
\Sigma_d = (-91.3 + 138.8) \; \; {\rm MeV} \hspace{1cm} {\rm (solution \; A)} 
\nonumber \\
\Sigma_d = (-94.5 + 144.2) \; \; {\rm MeV} \hspace{1cm} {\rm (solution \; B)}
\end{eqnarray}
where the first figure corresponds to the $d^+_{00}$ contribution and the
second the $2 d^+_{01}$ piece.

\section*{RESULTS AND DISCUSSION}

The problem with the sigma values used to be that the value for
$y$, the strange quark content, tended to be large in view of what could
reasonably be expected for the strange quark contribution for the
proton mass. Then it turned out [8,12], however,  that the $t$-dependence of
the scalar form factor was underestimated in the previous analyses, and
that the new value $\Delta_\sigma \simeq 15 \; \; {\rm MeV}$ led to a
more reasonable result, about 130 MeV of the proton mass coming from the
strange quark piece [12].

This analysis relied heavily on the Karlsruhe amplitudes, which 
included data from the 70's and earlier. So only a very limited amount of
information from the meson factories was incorporated. Especially at
low energies new measurements have been performed for different observables.
New data point to some changes in the Karlsruhe amplitudes, but there
is, at present, no general agreement of the new amplitudes, because
even the new data contain conflicts. Also, the discussion of the value of the
$\pi N$ coupling constant has continued with increasing vigour. In [12]
it was not possible to check the dependence of the values of $\Sigma$ on the
pion-nucleon coupling, because one was bound to use the coupling strength
corresponding to the input amplitudes, i.e. the Karlsruhe value 
$g^2/4 \pi=14.28$; $f^2=0.079$. The VPI group has, however, started to 
incorporate fixed-$t$ constraints to the phase shift analysis, which gives a
possibility to make some checks of the influence of the value of the
coupling. Their value for the $\pi N$ coupling is $f^2=0.076$ [14].  
Exploratory searches were reported in [17,18] for the VPI amplitudes
FA93 and SM95. Definite values for $\Sigma$ cannot be given, because the
curvature term $\Delta_D$ has not been estimated with these amplitudes, but
the linear part $\Sigma_d$ gets values of about 3 MeV larger than the ones
resulting from the use of the Karlsruhe amplitudes when the same low-energy 
data sets 
have been used as input. Another matter is that the favoured data sets would be
different from the ones which are consistent with the Karlsruhe amplitudes.
In any case, the effect of the $\pi N$ coupling seems to be relatively
unimportant in comparison with the error bar due to other uncertainties.
The reason could be that the $\Sigma$ involves the $\bar{D}^+$-amplitude where
the Born term has been subtracted and, therefore, the large contribution 
directly proportional to $f^2$ is removed from the amplitude. In [18] some
issues in the selection of the low-energy data was investigated for the
case of VPI-SM95 [14] input amplitudes. The trend seems to be towards slightly
larger values of the $\Sigma_d$, but the effect is typically about half
of the estimated total uncertainty, so the overall picture does not change 
essentially. 

New low-energy data has also appeared [19,20]. The differential cross
section data of Ref. [19] are
slightly too high in energy to be suitable for the present analysis. Ref.
[20] gives the first analyzing power results at low energy, and with more
data to come [21] an important constraint on the low-energy analysis will
be obtained. A careful discussion of the $\pi^+ p$ data base has recently
appeared in [22]. 

Another piece of new information has appeared quite recently, namely,
a measurement of the backward elastic $\pi^- p$ cross section [23] as a 
function of energy. With the KH80 input a good fit can be found, and the
result is $\Sigma_d = (-86.0 + 145.8) \; \; {\rm MeV}$. This result is
quite sensitive to the amplitudes and the P$_{33}$ splitting of
$\pi^- p$ and $\pi^+ p$. Even the $d$-waves start to matter, because the
VPI-SM97 $d$-waves differ upto a factor of two from the KA85 results.
No attempt was made to try to estimate the
error bar due to the errors in data.
A nice feature is, however, that the resulting
scattering lengths are agreeing quite well with the results
from pionic hydrogen experiment. The search gives
\begin{eqnarray}
a_{\pi^- p} & = & 0.0882 \; \; \mu^{-1} \nonumber \\
a^-_{0+}    & = & 0.0922 \; \; \mu^{-1}
\end{eqnarray}
to be compared with the experimental results from pionic hydrogen [24,25]
\begin{eqnarray}
a_{\pi^- p} & = & 0.0883 \pm 0.0008  \; \; \mu^{-1} \nonumber \\
a^-_{0+}    & = & 0.0920 \pm 0.0042  \; \; \mu^{-1}.
\end{eqnarray}

Another approach for the sigma-term discussion is gradually becoming more and
more useful. Namely, already now lattice calculations can make statements
about the value of sigma [26,27]. Using the Feynman-Hellmann theorem
\begin{eqnarray}
\sigma = \hat{m} \, \frac{\partial m}{\partial \hat{m}}
\end{eqnarray}
the sigma can be calculated from the quark mass dependence of the
proton mass. Of course, there are the
difficulties related to the question of dynamical quarks which until
now cannot be treated. Also, the quark mass values in actual calculations
are relatively large, but gradual improvement can be expected here. The
current numbers for $\sigma$ are 40-60 MeV [26], and 50 MeV [27] where
the latter calculation cites a very small error (3 MeV).

\section*{CONCLUSIONS}

In this workshop views were expressed that the modern $\pi N$
data base is essentially internally consistent. In the strongly constrained
low-energy analysis discussed here the picture is not quite that clear.
Also, there seems to be data sets which simply have to be ``thrown away''
in all analyses. In my opinion, one should aim at understanding from the
experimental point of view why there is so much trouble with the $\pi^+ p$
data. A good example is the disagreement of the differential cross section
results with the integrated cross section measurements [28,29].

Another set of problems is opened by the possibility of isospin violation
which has not been addressed here. The sensitivity of the extrapolation to the
Cheng-Dashen point may call for an improvement of the precision in this 
respect.

\subsection*{Acknowledgements}

I would like to thank R.A. Arndt for making SM97 available for me. In addition,
I thank J. Gasser and A.M. Green for useful remarks on the manuscript.

\bibliographystyle{unsrt}

\begin{thebibliography}{99}
\bibitem{1} J.~Gasser, and H.~Leutwyler, ``Quark masses,'' Phys. Reports
{\bf 87}, 77 (1982).
\bibitem{2} J.~Gasser, ``Hadron masses and the sigma commutator in light of chiral perturbation theory,'' Ann. Phys. (N.Y.) {\bf 136}, 62 (1981).
\bibitem{3} B.~Borasoy, and Ulf-G.~Mei\ss ner, ``Chiral expansion of baryon 
masses and $\sigma$-terms,'' Ann. Phys. (N.Y.) {\bf 254}, 192 (1997).
\bibitem{4} L.S.~Brown, W.J.~Pardee, and R.D.~Peccei, ``Adler-Weisberger 
theorem reexamined,'' Phys. Rev. {\bf D4}, 2801 (1971).
\bibitem{5} J.~Gasser, M.E.~Sainio, and A. \v{S}varc, ``Nucleons with chiral 
loops,'' Nucl. Phys. {\bf B307},779 (1988).
\bibitem{6} V.~Bernard, N.~Kaiser, and Ulf-G.~Mei\ss ner, ``On the analysis 
of the pion-nucleon $\sigma$-term: The size of the remainder at the 
Cheng-Dashen point,'' Phys. Lett. {\bf B389}, 144 (1996).
\bibitem{7} H.~Pagels, and W.J.~Pardee, ``Nonanalytic behavior of the $\Sigma$ term in $\pi$-N scattering,'' Phys. Rev. {\bf D4}, 3335 (1971).
\bibitem{8} J.~Gasser, H.~Leutwyler, and M.E.~Sainio, ``Form factor of the
$\sigma$-term,'' Phys. Lett. {\bf B253}, 260 (1991).
\bibitem{9} V.~Bernard, N.~Kaiser, and Ulf-G.~Mei\ss ner, ``Critical analysis 
of  baryon masses and sigma-terms in heavy baryon chiral perturbation theory,''
Z. Phys. {\bf C60}, 111 (1993).
\bibitem{10} R.~Koch, ``A new determination of the $\pi N$ sigma term using 
hyperbolic dispersion relations in the $(\nu^2,t)$ plane,'' Z. Phys.
{\bf C15}, 161 (1982).
\bibitem{11} J.~Gasser, H.~Leutwyler, M.P.~Locher, and M.E.~Sainio, 
``Extracting the pion-nucleon sigma-term from data,'' Phys. Lett. 
{\bf B213}, 85 (1988).
\bibitem{12} J.~Gasser, H.~Leutwyler, and M.E.~Sainio, ``Sigma-term update,''
Phys. Lett. {\bf B253}, 252 (1991).
\bibitem{13} R.~Koch, ``A calculation of low-energy $\pi$N partial waves 
based on fixed-$t$ analyticity,'' Nucl. Phys. {\bf A448}, 707 (1986).
\bibitem{14} R.A.~Arndt, I.I.~Strakovsky, R.L.~Workman, and M.M.~Pavan,
``Updated analysis of $\pi N$ elastic scattering data to 2.1 GeV: The baryon
spectrum,'' Phys. Rev. {\bf C52}, 2120 (1995).
\bibitem{15} G.~H\"ohler, {\em Landolt-B\"ornstein, Vol. 9 b2} 
(Springer, Berlin, 1983).
\bibitem{16} B.~Tromborg, S.~Waldenstr\o m, and I.~\O verb\o,``Electromagnetic
corrections to $\pi N$ scattering,''
Phys. Rev. {\bf D15}, 725 (1977).
\bibitem{17} M.E.~Sainio, ``Pion-Nucleon Sigma Term,'' in {\em Proceedings
of the Workshop on Chiral Dynamics: Theory and Experiment} (Springer, Berlin
and Heidelberg, 1995) pp. 212-223.
\bibitem{18} M.E.~Sainio, ``Update of the sigma-term,'' $\pi$N Newsletter
{\bf 10}, 13 (1995).
\bibitem{19} J.T.~Brack et al., ``Forward angle $\pi^\pm p$ elastic scattering
differential cross sections at \mbox{T$_\pi$ = 87} to 139 MeV,'' Phys. Rev.
{\bf C51}, 929 (1995).
\bibitem{20} R.~Wieser et al., ``Measurement of the $\pi^+\vec{p}$ analyzing
power at 68.3 MeV,'' Phys. Rev. {\bf C54}, 1930 (1996).
\bibitem{21} G.J.~Hofman, ``$\pi$P analyzing powers with the CHAOS 
spectrometer,'' $\pi$N Newsletter {\bf 10}, 128 (1995).
\bibitem{22} N.~Fettes, and E.~Matsinos, ``Analysis of recent $\pi^+ p$
low-energy differential cross-section measurements,'' Phys. Rev.
{\bf C55}, 464 (1997).
\bibitem{23} M.~Janousch et al., ``Destructive interference of s and p waves
in 180$^\circ$ $\pi^- p$ elastic scattering,'' preprint ETHZ-IPP PR-97-03 
(1997).
\bibitem{24} D.~Sigg et al., ``The strong interaction shift and width of
the ground state of pionic hydrogen,'' Nucl. Phys. {\bf A609}, 269 (1996);
(E) Nucl. Phys. {\bf A617}, 526 (1997).
\bibitem{25} M.~Janousch et al., ``Determination of the $\pi$N s-wave 
scattering lengths in pionic hydrogen and deuterium,'' in {\em Proceedings
of the 14th International Conference on Particles and Nuclei} 
(World Scientific, New Jersey, 1997) pp. 372-373.
\bibitem{26} M.~Fukugita, Y.~Kuramashi, M.~Okawa, and A.~Ukawa, 
``Pion-nucleon $\sigma$ term in lattice QCD,'' Phys. Rev. {\bf D51}, 5319 
(1995).
\bibitem{27} S.J.~Dong, J.-F.~Laga\"e, and K.F.~Liu, ``$\pi$N $\sigma$ term,
$\bar{s}s$ in the nucleon, and the scalar form factor: A lattice study,''
Phys. Rev. {\bf D54}, 5496 (1996).
\bibitem{28} B.J.~Kriss et al., ``Pion-proton integral cross section 
measurements,'' $\pi$N Newsletter {\bf 12}, 20 (1997).
\bibitem{29} E.~Friedman et al., ``Integral cross sections for $\pi^+$p
interactions at low energies,'' Nucl. Phys. {\bf A514}, 601 (1990) and
private communication.

\end{thebibliography}

\end{document}